\documentclass{IEEEtran}

\usepackage{url}
\usepackage{tikz}
\usetikzlibrary{arrows}
\usetikzlibrary{backgrounds}
\usepackage{enumitem}

\begin{document}
\title{Digital Encycplopedia of Scientific Results}
\author{J\'anos Tapolcai}
\maketitle

\begin{abstract}
This study describes a vision, how technology can help improving the efficiency in research. We propose a new clean-slate design, where more emphasis is given on the correctness and up-to-dateness of the scientific results, it is more open to new ideas and better utilize the research efforts worldwide by providing personalized interface for every researcher. The key idea is to reveal the structure and connections of the problems solved in the scientific studies. We will build the system with the main focus on the solved problems itself, and treat the studies only as one presentation form.   
By utilizing artificial intelligence and machine learning on the network of the solved problems we could coordinate individual research activities in a large scale, that has never been seen before.
\end{abstract}

\section{Introduction}

Research is about searching for answers on very difficult questions, e.g. 
\emph{How did climate changed in the recent 30 years? How can cloud technologies change data storing? How can we cure AIDS? etc.}
The full answers to these questions are way too complex to be understood by a single human; thus, these problems are cut into small parts that can be answered by a group of people. We call these small parts as \emph{sub-problems}, and with solutions we treat them as the main building blocks of  science. Eventually, the final answer to the very difficult questions is going to be a joint work of many researchers working independently, sometimes even in different centuries. The only glue that connects these researchers are text documents in which researchers write down their new findings. We call these documents \emph{studies}, and they may have various formats (such as papers, theses, reports and books) but overall they follow roughly the same structure since many centuries.  

There have been many efforts in building up digital databases over these studies, such as ScienceDirect, Scopus, Google Scholar, CiteSeer\textsuperscript{x}, Microsoft Academic Search, etc \cite{ortega2014academic}. They provide advanced search and keep track the citation network, which greatly improved the efficiency of research in the last decade. The key statement of the current study is that the time is ripe: technology is ready to provide a way more efficient tool for researchers after a clean-slate design. The key idea is to built the system around the scientific sub-problems and their relationships. Currently the main focus is on the scientific studies, and the structure of the sub-problems remains hidden behind them. Lunching such a system requires a significant amount of intellectual work by transforming existing scientific studies into a network of solved sub-problems. This study overviews how this can be done and what can be the incentives.

%

\section{State of the Art}

Before we further discuss our goals let us overview the typical properties and possible weaknesses of the current system. It considers research studies as the central element that follows the unique style and logic of each research field. These styles can be quite different, but the following building blocks appear in all of them with different emphasis:
\begin{description}
\item[sub-problems:] the research questions addressed in the study. It is often in the form of problem definition  or theorems.
\item[solutions:] the answers to the above questions (also called the main contributions). 
\item[positioning:] the relationship of the study compared to previous studies. 
For example, how the addressed sub-problem is related to more general research questions. 
\item[supporting material:] which are summaries (e.g. abstract and conclusions), and all type of extra information about the work (e.g. motivation, background of the work, list of authors, acknowledgements, etc).  
\end{description} 
An important property is that once a study is \emph{published} it remains as is forever, and cannot be edited anymore. We have seen many examples where the development of a research field was temporary stopped because of some badly written section in an important study. For example, in mathematical papers it is common to ignore some parts of the proof by saying ``one can easily see". Even if the paper is brilliant, 
later it might not be trivial for a reader with different background than the authors, and make the solution incomplete. On the other hand, publishing these missing parts of the proofs as an independent study is rarely done as it is considered to be a very incremental contribution. The extended proof can be included in a book overviewing the whole research field, but this is also a huge work. 

If an error is found in a study it is possible to publish an \emph{errata} to correct it. Withdrawing a study is also possible, but it is very rare and considered to be a disgraceful event. Before a study is published it is almost always peer reviewed in ordered to verify it by a few anonym researchers. The reviewers often have a chance to suggest some changes in the study before it is published. Publishers of the prestigious venues pay special attention to have high quality reviews, and be able to select the most important studies for publishing. Many pointed out the weak points of peer review systems, but so far there is no successful alternatives. For example 
usually the reviewers do not get any credit, and thus it is always hard to convince (and find) the most proper reviewer. 

The publishing venues (e.g. journal, conference, book) have several requirements in order to make the work more readable for humans. For example length of the published study is defined by a strict page-limit. 
Another typical human constraint is sanctioning plagiarism in academia. Each study should avoid using another author's language, thoughts, ideas, or expressions. 
Although these constraints are all reasonable, but sometimes have negative influence on how research evolves.
We argue that \emph{currently too much emphasis is given on the presentation of research results, and it is far from the optimal structure.} 

\section{Goals}

\subsection*{Goal 1: Maintain an editable map of sub-problems}

We treat research as a process where a great number of sub-problems are solved to eventually find the answer on a question addressed. In our vision besides the solutions to sub-problems, the relationships of these sub-problems are also of central importance, which we call \emph{the map of the sub-problems}. This map is changing as the research evolves and its details are often hidden in the studies. For example a study deals with two sub-problems $A$ and $B$, where $B$ is a special case of $A$ (any solution to sub-problem $B$ is also a solution to sub-problem $A$, while the opposite is not true). 

We take a special care of the citations. First we allow to precisely define which sections of the cited paper they are referring to. Many studies cover several sub-problems and it is not simple to map the particular sub-problems cited in a paper.  Second, we list the paragraphs/sentences how a study is cited later in the literature. This allows to navigate forward in the citation graph to find a latest result related to each sub-problem. 

\subsection*{Goal 2: Emphasis the links between sub-problem (and papers)}

The map of sub-problem can be treated as a network, where the nodes are the sub-problems and the links are the relationship between the sub-problems. To understand this network we need to discover all possible links between the nodes. Our motto is that  ``the main contribution is often lays not within the paper, but between the papers''. Unfortunately, the current structure of research does not favor discovering these links, but rather focuses on the nodes. For example it is very rare that the only contribution of a study is to show that a sub-problem in paper $A$ is the same as the sub-problem of paper $B$. This type of contribution is often considered incremental, and thus easily becomes forgotten by not receiving sufficient highlights. At the same time research activity is greatly increased in the last decades, and many small research communities were born focusing on one line of research evolving from a sub-problem. These communities are large enough to progress in one direction, and to make them believe  other research communities are working on totally different sub-problems. Nevertheless, we strongly believe there is significant overlapping between them, and ``reinvention of the wheel in research'' is a factor that can be reduced.  
   
\subsection*{Goal 3: Lower the entry threshold for research}
 
Researchers in each field form a narrow pyramid where the top researchers have huge influence, while it is negligible small at the bottom of the pyramid (e.g. student from less economically developed countries). Luckily, great talents usually find the way to get higher in the pyramid. Nevertheless, we believe a wider and lower pyramid would make research way more efficient. The success of open source software implementations proved how computer systems can efficiently organize collaborative working. 
We plan to manage a large database on the background knowledge of each researcher. 

Pioneering results by top researchers often have effects on a great number of sub-problems. If we can identify these small possible contributions by analyzing the network of subproblems, and able to implement a strong validation process, then we could suggest them for students to work on, and get them involved in top research as well. 
In particular, by running machine learning algorithms in the network of subproblems, we should be able to identify interesting hotspots and assign for a suitable candidate. Technically, a user can request the system to point on a small set of papers that has a high potential to inspire new contributions. The user later can submit a short discussion how the sub-problems are related to each other.

An interesting successful example of an on-line collaborative research platform is the ``Polymath project'' created by Terence Tao of the University of California, Los Angeles, a winner of the Fields Medal, mathematics' highest honor. Here, a massive on-line collaboration of a dozen of researchers have made enormous advances over six months on a centuries-old twin primes conjecture \cite{polymath}.
\subsection*{Goal 4: Keep the unconstrained structure of research}

We will keep written English text (optionally with equations and figures) as the main way of presenting results. Written text is an unconstrained form of expressing ideas. We also allow to add ``translations'' of the sub-problem using different terminology and notation.  To design the system very flexible, we will not define any research fields, and instead store as many translations of the same sub-problem as possible (even if it is just a replacement of one world to an other, like ``nodes'' to ``vertices'' in graph theory). 

Technically, we intend to use latex where the notations are defined by macros that enables to simply adapt the text to different terminology. The idea is that when a text object is cited we have the opportunity to re-define the  notation and terminology. In this way a cited object will appear using the same notation and terminology as in the original paper. The confusion in the terminology between paper is often the source of errors in research. 

We also allow adding any other digital contents like: presentation slides, videos, source codes, and all types of data, etc.  

\subsection*{Goal 5: Ensure the validity of the research results}

We should avoid that fake results appear in research and in our database, thus every objects (questions, answers, translations, links, etc.) should be verified by reviewers. Instead of traditional peer review systems we will propose a different mechanism, where reviews are simpler (yes or no) and the identity of the reviewers can be revealed. We expect this will significantly increase the number of reviews, and make the results more solid. The name of a famous scholar who verified the object is the best certificate it can get. In our view, an idea is not necessarily finished when it is ``published'', but it evolves as more and more researchers understand and verify its correctness. If a reviewer finds some parts confusing, he/she can add a translation, which will give credits for both the authors and reviewer. The reviewer can identify missing parts of the work (as new sub-problems), which can be solved by the original authors or by someone else. 

We plan to publish a quantified metric on each contributors reviewer's activity, to give credits for their work. We also plan to keep track the responsibility of the contributors. The authors will be motivated to submit mature and correct work, to have high contributor index. The contributor and reviewer index will define a pyramid of the researchers, and for each object we will try to automatically assign some reviewers that are higher in the pyramid than the author. In this way if a newcomers submits a work, it will receive a review from the bottom of the pyramid first, etc. 


We also plan to list how each object was cited in the later studies. It can help identifying the possible problems and limitations of the solutions, which were not discovered (or presented) by the authors. For example many study have numerical evaluations to support the main messages. However, it is very hard for the reader to judge wether the obtained measurements are general, or they were carefully selected to best support the contributions, while some of the measurements (not supporting the claims) were simply ignored. This issue, call ed reproducibility, got quite some attention recently. 

\subsection*{Goal 6: Provide methods to evaluate the contributors}

Currently there are only very rough metrics available to quantify researchers. Most of them corresponds citation analysis (like, independent citations, H-index, etc). These metrics suffer from many known weaknesses when credits are not given where credit is due, like secondary sources where review articles collect the credit instead of the original sources, etc. 
By understanding the structure of research results, we can provide more accurate methods to evaluate the author's contributions, numbers that even non-professionals can understand.  This can help in making ``unbiased'' decisions concerning grants, promotions, etc. 

\section{Proposed system design}

A Wikipedia style online webpage storing the editable map of sub-problems targeted in research papers.  Solutions can be added to these sub-problems, and studies can be build up from these objects along with the supporting material.
We also introduce several new objects, like
\begin{itemize}
\item Links between problems, for example: a solution to Problem $A$ is a solution to Problem $B$ as well.
\item Objects to better specify a reference with explanation how to transfer notations, and the exact sentences referring to.
\item Add a new property of any set of objects e.g. add a new keyword or hashtag.
\item Add labels to an existing object, with remarks, notes, errata, translation. 
\item Allow discussions between contributors by messages and on-line conversations.
\end{itemize} 

Based on the above database the following potential benefits we expect:
\begin{description}
\item[Easier to learn a field] Wikipedia has showed how efficient can be to learn a new field when the text is extended with hyperlinks. When a page is linked we also define the notation and terminology transformation, so when it is loaded the notation is inline with the previous page. This helps to reduce the confusion caused if the same notation is used for two different things on different pages.
\item[Easier to find results] Be knowing the relationship of the sub-problems, will will be able to implement a very advanced search among the research results.
\item[Introduce Artificial Intelligence in research] Running machine learning on network of subproblems we can automatically identify potential similarities and the most important issues that need sot be addressed. It can automatically discovering all the consequences of a new solutions. By understanding how each sub-problem relates to fundamental research problem, we may be able to shepherd researchers to work on topics most important for the society.  
\item[Personalized interface for researchers] The system can identify the background knowledge and style of  each contributor. Based on this the system will be able the suggest ``interesting'' objects (links) to validate. It can also suggest interesting sub-problems or (recent) solutions of high importance for researcher. It can initiate a  conversation between different researchers on a specific topic by knowing their background and interest.
\end{description}

\section{Risk management}

\subsection*{Difficulty 1: There is no common language}
Research papers are mostly written in English, however each community has their very own style. Being a novice in a research area it is considered to be a challenging task to read the first paper, even if you have a lot of experience in other research fields. It is because the paper structure and logic, the terminology and notation, the format, and usage of figures are all varies among the fields. Papers in each area shortens (or leaves out) the part which is trivial (was already said many times before) or very technical or non-relevant. These parts can be challenging to reproduce for someone with different background, e.g. in engineering papers the targeted sub-problems are often explained through simple examples, and a formal definition is omitted. As a solution we will encourage translating sub-problems to different styles, in order to help identifying the papers with similar topics in different fields. Currently, such translations are not a recognized scientific value, and thus it is rarely done by experts.

Having such translations, we can show the sub-problems in the translation closest to the reader's style. We plan to adapt a latex style interface, where the notations of a study are defined by macros and can be easily changed.

\subsection*{Difficulty 2: It conflicts with the current business model in research}

Research is mostly funded by governments, mainly as a part of common wealth. It paybacks in short term if research can be transformed into innovation. However, usually research is a long term investment, and countries investing in knowledge are more successful. There is also some targeted industrial support, but not with the aim to publish the great ideas.
Governments fund research through selected projects and researchers, which generates a staggeringly profitable business of scientific publishing.
Overall there is room to improve the efficiency of current research funding systems.
  
In short term we treat our system as an extension to the current model. However, we see the value of conferences in long term. Conferences can summarize and highlight recent results, and are great opportunities  for researchers to meet. Our plan to dedicate conferences where the studies with highest impact are presented each year. 

In long term our gaol is to convince funding agencies to support research through our system. Based on the network of results we can better quantify the contributions of each researcher. We can also allow to address questions in the system and assign a price for it, which is divided between the many contributors in fair way. Addressing questions is also open for companies, which can provide them an affordable interface for researchers, while researchers can work on with real life problems.
Nevertheless, we need continuous income to keep the system up to date.

\subsection*{Difficulty 3: How to validate research results?}

Validating the results is the key of the success of the system.  It is in general very challenging, as we have seen many groundbreaking contributions which were not appreciated when they first come out. For this purpose, 
if the validation is controversial, we will initiate a discussions during the validation between the reviewers, and keep it in the loop until a consensus is found.

As we have discussed earlier for each object we will show the list of contributors who validated.  It can be done with an interface similar to Stack Overflow, where the best solutions can be upvoted. We will keep track of the level of confidence of each contributor.  Some of the budget can be allocated for reviewing purposes. 
We also pan to use artificial intelligence to identify malicious researchers \emph{not} collaborating. 


\section{An illustrative example}

\begin{figure}
\tikzstyle{question}=[draw, fill=blue!2, minimum size=2em,font=\footnotesize]
\tikzstyle{question2}=[question, text width=1.2cm,font=\scriptsize,node distance=1.4cm]
\tikzstyle{init} = [pin edge={to-,thin,black}]

\begin{tikzpicture}[node distance=1.5cm,auto,>=latex',font=\footnotesize]
    \node [question] (a) {\textbf{Problem 1:} How to schedule jobs in data center?};
    \node [question, below of = a] (b) {\textbf{Prob. 2:} Prob. 1 with heterogenous  computers and single resource};
       \path[->] (a) edge node {special case} (b);
    \node [question, below of = b, text width=8cm] (c) {\textbf{Prob. 3:} Prob. 2 meeting constrained sharing incentive and strategy-proof properties};
       \path[->] (b) edge node {special case} (c);
     \node [question, below of = c] (d) {\textbf{Prob. 4:} Constrained Max-Min Fairness (CMMF) problem};
       \path[->] (c) edge node {special case (Thm. 2 and 3)} (d);
      \node [question2, below of = d, xshift=-35mm, yshift=0mm] 
                                                       (e1) {\textbf{Prob. 5:} offline, \\divisible, weighted CMMF}; 
      \node [question2, right of =e1] (e2) {\textbf{Prob. 6:} offline, \\divisible, unweighted CMMF};  
      \node [question2, right of = e2] (e3) {\textbf{Prob. 7:} offline, \\nondivisible, weighted CMMF}; 
      \node [question2, right of = e3] (e4) {\textbf{Prob. 8:} offline, \\nondivisible, unweighted CMMF}; 
      \node [question2, below of = e3,yshift=-5mm] (e5) {\textbf{Prob. 9:} online, \\divisible, weighted CMMF}; 
      \node [question2, right of =e5] (e6) {\textbf{Prob. 10:} online, \\divisible, unweighted CMMF};  
      \node [question2, right of = e6] (e7) {\textbf{Prob. 11:} online, \\nondivisible, weighted CMMF}; 
      \node [question2, right of = e7] (e8) {\textbf{Prob. 12:} online, \\nondivisible, unweighted CMMF};        \begin{scope}[on background layer]
          \draw[->] (d) -| (e1);
          \draw[->] (d) -| (e2);
          \draw[->] (d) -|  (e3);
           \draw[->] (d) -|  (e4);
           \draw[->] (d) -| (e5);
           \draw[->] (d) -| (e6);
           \draw[->] (d) -|  (e7);
          \draw[->] (d) -|  (e8); 
       \end{scope}
\end{tikzpicture}
\caption{\label{fig:choosy} The sub-problems and their relationship of study \cite{ghodsi2013choosy}.}
\end{figure}
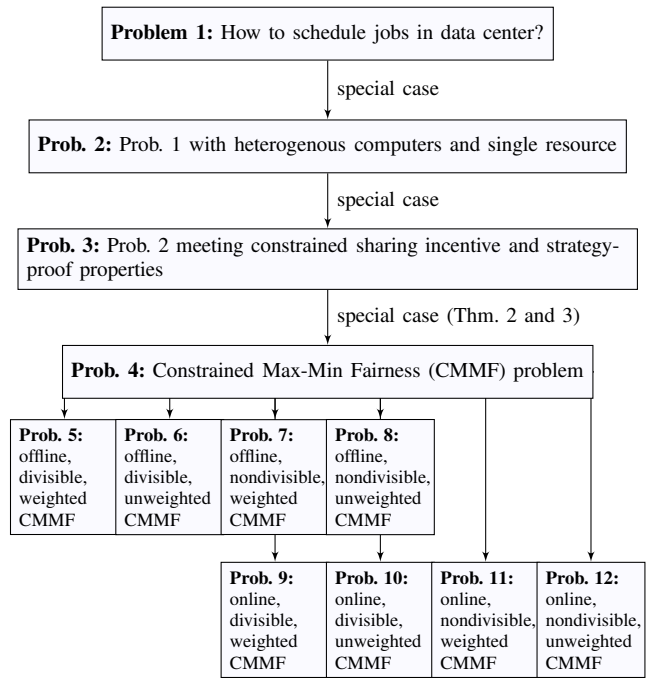

In this section we will give an example by examining the paper "Choosy: Max-Min Fair Sharing for Datacenter Jobs with Constraints" 
by Ghodsi, Zaharia, Shenker and Stoica \cite{ghodsi2013choosy}. 
The paper focuses on a sub-problem of the general question \emph{How to schedule jobs in data center?} The sub-problems and their relationship of the study is shown on  Fig. \ref{fig:choosy}.

Based on the above sub-problem structure we can assign each subsection to one of the boxes or arrows. For example Sec. 2 provides an example and explains the background of Prob. 2. Sec. 3 provides a discussion about Prob. 3. 
Thm. 2 and 3 shows why CMMF meets the constrained sharing incentive and strategy-proofness properties. The example of Fig 3 and 4. explains the CMMF problem. Sec. 5 deals with the algorithmic solution to Prob. 5-8, Sec. 6 with Prob.  9-12.  Finally, Sec. 7  provides an evaluation of the solutions given to Prob. 5-12.

Sec. 8 corresponds to Related Works, which need to be split between the above problems, but mostly they correspond to Problems 1 and 2. Usually, the related works section is mainly written for reviewers with the aim to highlight the differences compared to the previous works, and to show that the results are original. Therefore the authors are motivated to find the differences instead of discovering the similarities, which can be an important information for future readers. We suggest to add each reference (or group of references) as separate link objects between the corresponding sub-problems with the text describing the relationship. This will allow later to add remarks to these link object if similarities are found. If the cited paper is not part of the database we can add it as one study object, and later the study object can be still divided into sub-problems.   

We also define a \emph{study} object for this paper as well, which defines the structure of the paper. It includes the title, authors, abstract, introduction section, reference to the Problems 2-12 and solution objects and finally the conclusions. The aim of the study object is to provide a ``linear'' description of the work, where the order and the emphasis of each section is optimized for easier understanding. 
Nevertheless, the presentation of the paper is excellent for researchers in the networking field, it is way too vague for someone with computer science background because of the lack of formal mathematical problem definitions. 

When formalizing the problems one  may notice that Prob. 8 is equivalent to a hyper-edge orientation problem in hyper-graphs with a cost function of having lexicographically-maximal (egalitarian) node degrees (e.g. \cite{borradaile2012egalitarian}). 
It would be interesting to add such contribution to the paper which would provide a link to matroid theory. 
Note that, single-source un-splittable flow problem can be regarded as a matroid. It is also mentioned under scheduling theory in the related work section, and the relationship between the two problems is described with the following sentence: ``This work differs from ours in that we are concerned with parallel jobs that are composed of multiple tasks and can thus be assigned multiple machines, as opposed to the unsplittable jobs in Kleinberg et al.'s work.''  In the works of Kleinberg et al. \cite{kleinberg1996single,kleinberg1999fairness} the terms ``parallel'' or ``multiple tasks''  are not used, which makes confusion in the reader. The references would benefit from clear transformation between the terminology of the two papers. In my understanding Kleinberg et al.'s algorithms solves Prob. 8.    
Note that, network flows can be splittable or unsplittable in the same way as the jobs are either divisible or nondivisible in the paper. These issues worth clarification, and adding them as remarks to the link objects would help the reader better understanding the work.

Discovering the relationship between matroid (network flow) theory and the CMMF problem could be an interesting work for a master student in computer science. The greedy algorithm solving the related matroid (intersection) problem would be a new algorithm compared to the ones in the paper.  Moreover the general question of identifying the connections between efficient algorithms (greedy) and cloud job scheduling could be an important contribution for both fields. However, first, it requires to translating the related papers to both languages that engineers an computer scientists can read.

\bibliographystyle{IEEEtran}
\bibliography{ref}

\end{document}